%% file: Main.tex
\documentclass[a4paper]{article}
\usepackage{soul, color}
\usepackage{parskip}
\usepackage{amssymb}
\usepackage{outlines}
\usepackage{graphicx,subfig,subcaption}
\usepackage{amsmath,relsize}
\usepackage{ragged2e}
\usepackage[T1]{fontenc}

\usepackage[margin=1in]{geometry}

\usepackage{newfloat}
\DeclareFloatingEnvironment[
  fileext = lofa,
  placement = htbp,
  name = Figure,
  within = section
  ]{figureSupp}
  
\DeclareFloatingEnvironment[
  fileext = lofa,
  placement = htbp,
  name = Table,
  within = section
  ]{tableSupp}
\usepackage{caption}
\usepackage[title]{appendix}

\begin{document}

\title{Deterministic Detection of Single Ion Implantation}

\maketitle

{\large
\author{Mason Adshead*,}
\author{Lok Kan Wan,}
\author{Maddison Coke,}
\author{and Richard J Curry}
}


Department of Electrical and Electronic Engineering, Photon Science Institute, University of Manchester, Manchester, M13 9PL, UK\\
\\
*Email Address: mason.adshead@manchester.ac.uk
\\

\vspace{30pt}

\justifying
\begin{abstract}

\input{Sections/Abstract}
\end{abstract}

\input{Sections/Intro}

\input{Sections/DetectionEfficiency}

\input{Sections/ResultsAndDiscussion}

\input{Sections/SummaryAndConclusions}

\input{Sections/Method}

\medskip

\textbf{Supporting Information} \par
Supporting Information supporting the work presented in this publication is available from the author.

\medskip
\textbf{Acknowledgements} \par
This work was supported by funding from the UK EPSRC grant EP/V001914/1.

\medskip

\pagebreak
\bibliographystyle{MSP}
\bibliography{Sections/refs}

\pagebreak


\end{document}

%% file: Sections/Abstract.tex
Single ion implantation using focused ion beam systems enables high spatial resolution and maskless doping for rapid and scalable engineering of materials for quantum technologies, particularly qubits and colour centres in solid-state hosts. In such applications, the confidence with which a single ion can be deterministically  implanted is crucial, and so the efficiency of the detection mechanism is a vital parameter. Here, we present a study of the single-ion detection efficiency for a variety of ion species (Si, P, Mn, Co, Ge, Sb, Au and Bi) into various hosts (Si, SiO\textsubscript{2}, Al\textsubscript{2}O\textsubscript{3}, GaAs, diamond and SiC). The effect of varying ion mass, charge and kinetic energy are studied, in addition to the cluster implantation of Sb, Au and Bi. We demonstrate that it is possible to achieve detection efficiencies $>$90\% for a wide range of ion species and substrate combination through selection of the implantation parameters. Furthermore, detection efficiencies of 100\% are found for the doping of Sb clusters which is of direct relevance for the future fabrication of quantum devices.

%% file: Sections/Intro.tex
\section{Introduction}

The realisation of robust methods for fabricating buried donor-ion solid-state qubit technologies is essential to enable their scaling into large arrays to realise fault-tolerant quantum computing. Large numbers ($\sim$ 1 million) of physical qubits are required in order to employ quantum error correction protocols, and to deliver the number of logical qubits required to ensure quantum supremacy \cite{kim2024fault}. Achieving this capability within solid-state hosts requires the placement of dopants with high spatial precision and low doping error rates, meaning large-area doping methods such as gas-assisted deposition and doping during substrate growth are unsuitable. Lithographic methods based on scanning tunnelling microcopy have the necessary high precision placement of dopants, however they present challenges in scaling to deliver the very large arrays required and are subject to the stochastic nature of dopant formation, meaning that sometimes clusters of ions can form when only a single ion is desired \cite{wyrick2019atom}. Ion implantation as a versatile and scalable technique for solid-state doping is well-established. It has been successfully used to create physical qubits via low-dose broad beam doping using nano-apertures for high resolution spatial positioning \cite{jakob2024scalable}. Recent advances in focused ion beam (FIB) doping have enabled its application to demonstrate functionality aligned with the scalable delivery of quantum technologies \cite{adshead2023high}. These include localised isotopic enrichment of \textsuperscript{28}Si to provide nuclear-spin-free hosts for donor ion qubits \cite{acharya2024highly, coke2025fabrication} and fabrication of large arrays of single-photon emitters with high spatial positioning capability in diamond substrates \cite{cheng2025laser}. FIB implantation therefore provides an attractive methodology for the scalable doping of donor ions within solid-state materials for qubit formation.

A critical requirement for the formation of a working array of physical qubits is to control of the number of ions doped at each location. Multiple ions at the same location will result in undesirable ion-ion interactions leading to decoherence and impacting qubit control. A detailed study of the experimental factors that govern the deterministic doping of single ions is given by Murdin, et al. \cite{murdin2021error} with single-ion detection being a key requirement. The detection of single-ion implantation events in to solid-state substrates, such as diamond and silicon, can be performed using ion beam induced charge (IBIC) detection. This relies on the detection of secondary electrons generated within the material by the implanted ion, using an electric field to sweep the carriers to an electrode on the substrate surface (or bottom side). Single ion detection efficiencies using this approach have been reported of 85.5\% in graphene-enhanced diamond \cite{collins2023graphene}, 87.6\% \cite{pacheco2017ion} and  $>$99\% \cite{jakob2022deterministic} in Si-based detectors for ion implantation energies in the range of 10s of keV. However, despite the high efficiencies achieved, the requirement to fabricate in-built IBIC detectors within the target materials adds complexity and cost that often precludes them from being integrated and scaled to produce large array devices.

An alternative approach to single ion implantation detection utilises the secondary electrons ejected from the surface of a substrate as a consequence of ion implantation with high sensitivity secondary electron detectors (SEDs). This method of emitted secondary electron detection does not require any additional processing, e.g. electrode deposition to form an in-built device, and therefore offers a simple method applicable irrespective of the target substrate. Cassidy, et al. \cite{cassidy2021single} previously investigated the SED detection efficiency of Bi ion implantation into a small number of substrates, reporting detection efficiencies into silicon of up to $89$ $(\pm2)$ \% for 50 keV Bi ions.

In this work, we report a detailed study of the influence of ion species (mass), charge state and energy on single-ion and single-ion-cluster detection efficiency into a variety of substrates. We demonstrate that through choice of implanted ion, energy and target substrate detection efficiencies up to 100 \% may be achieved.

%% file: Sections/DetectionEfficiency.tex
\section{Deterministic Ion Implantation}

The determination of detection efficiency, $\eta$, for the current work is based on the method described by Cassidy, et al. \cite{cassidy2021single}. A detailed outline of the ion beam pulsing and detection configuration specific to this work is given in the Experimental Methods section.

For a series of ion beam pulses, the average number of detections per pulse, $\nu$, can be used to determine $\eta$:

\begin{equation}
    \nu = -\ln \left( \frac{M-p}{M} \right) = \eta \lambda
    \label{eq:nu+etaLambda}
\end{equation}

with $\lambda$ defined as

\begin{equation}
    \lambda = \frac{It}{q}
    \label{eq:lambda}
\end{equation}

where $p$ is the number of successful detections from $M$ attempted implantations. $\lambda$ is the average number of ions per pulse, with $I$ the ion beam current, $t$ the ion pulse width, and $q$ the ion beam species charge state (in multiples of the universal electronic charge, $e_c$).

In the work presented here, the ion beam current was fixed and $\lambda$ varied by changing the pulse width. The number of required positive detections for experiment completion, $p$, was maintained at 2500 (determined by implantation into a 50 x 50 point array as described in the Experimental Methods section). The number of pulses required to record the 2500 positive detections varied, depending on three main factors: (i) the average number of ions in a given pulse, $\lambda$; (ii) the emitted secondary electron yield for the ion implanted into a substrate (which varies based on the ion beam and substrate combination); and (iii) the efficiency of the detection mechanism.

\subsection{Linear Regression}

The determination of a successfully implanted ion is based on the SED signal correlated with the pulsing of the ion beam. Software control of the implantation experiments registers a signal above a noise threshold (set at 500 mV for this work) of the amplified signal as a detected implantation. The software analysis of the SED signal is binomial (for each ion pulse delivered an implantation event is either detected or not detected) and as $\nu$ scales its error also scales, therefore linear regression is appropriate in the determination of the error in the average number of detections per pulse, $\nu$ \cite{murdin2021error}. Ion beam blanking is achieved by the biasing of electrostatic plates, of 2 mm length, through which the beam passes. The bias is briefly removed for time $t$ to create an ion pulse. The transit time of ions through the 2 mm plates, $T_0$, results in a necessary adjustment to the pulse width used in calculations of $\lambda$.

\begin{table}[]
    \centering
    \begin{tabular}{c|c|c|c|c|c}
    Set Pulsewidth & $\lambda$ & Num. Attempts, & $\nu$ & $S(\nu t)$ & $S(t^2)$  \\
    ($\mu s$) & (ions/pulse) & $M$ & & \\
    \hline
    0.21    & 0.0953  & 36257 & 0.0769 & 6707 & 18000 \\
    0.42    & 0.1923  & 17245 & 0.1494 & 6954 & 19382 \\
    0.63    & 0.2893  & 11320 & 0.2430 & 6412 & 16526 \\
    0.84    & 0.3863  & 8655 & 0.3121 & 6645 & 17805 \\
    1.05    & 0.4833  & 7089 & 0.4043 & 6383 & 16518 \\
    1.26    & 0.5803  & 6175 & 0.4849 & 6352 & 16457 \\
    1.47    & 0.6773  & 5502 & 0.5465 & 6544 & 17560 \\
    1.68    & 0.7743  & 5148 & 0.6527 & 6197 & 15916 \\
    1.89    & 0.8713  & 4478 & 0.7536 & 5970 & 14945 \\
    2.20    & 0.9683  & 4334 & 0.8216 & 6032 & 15390 \\
    \end{tabular}
    \caption{Detection data for 50 keV P into intrinsic silicon with a native oxide layer, with an ion beam current of $148 \pm 7$ fA. The values for $s(\nu t)$ and $s(t^2)$ are determined using the linear regression method outlined by Murdin, et al \cite{murdin2021error}. The switching latency was calculated to be 4 ns, which was used to find the adjusted pulse width ($t=T-T_o$) for the calculation of $\lambda$.}
    \label{tab:exampleDetection}
\end{table}			

An example of detection efficiency data for 50 keV P (25 kV anode voltage, Wien filter selecting the doubly charged ion, P\textsuperscript{2+}) into silicon (with a native oxide layer) is shown in Table \ref{tab:exampleDetection}. The pulse width was varied for a fixed ion beam current ($148 \pm 7$ fA) and the number of pulse attempts to reach 2500 successful detections was recorded. The representative number of ions per \textmu s, $N$, with its associated error, $\delta N$ were obtained using:

\begin{subequations}
    \begin{equation}
        N =\sum \frac{s(\nu t)}{s(t^2)}
        \label{eq:N}
    \end{equation}
    \begin{equation}
        \delta N = \frac{1}{\sqrt{\sum s(t^2)}}
        \label{eq:dN}
    \end{equation}
\end{subequations}

where, for each iteration ($i$) of $\lambda$, the pulse length, $t$, and average detections, $\nu$, give:

\begin{subequations}
    \begin{equation}
        s(\nu t)=\sum_i\frac{\nu_i t_i}{s(\nu_i)^2}
    \end{equation}
    \begin{equation}
        s(t^2)=\sum_i\frac{t_i^2}{s(\nu_i)^2}
    \end{equation}
\end{subequations}

with the number of attempted pulses for a given iteration of $\lambda$, $M_i$, providing:

\begin{equation}
    s(\nu_i)=\sqrt{(e^{\nu_i}-1)/M_i}
\end{equation}

The detection efficiency, $\eta$, can then be determined using the expression:

\begin{equation}
    \eta =\frac{N}{L}
\end{equation}

where $L$ is the number of ions per \textmu s, determined using Equation \ref{eq:lambda}, with $t=1$ \textmu s. For the data presented in Table \ref{tab:exampleDetection} the detection efficiency is calculated to be $82$ ($\pm 4$) \%, the uncertainty for which comes from the error propagation of $L$ and $N$.

The detection efficiency calculated here is representative of the probability of successfully detecting an ion implantation event for a specific point. This can be extended to obtain the probability of successfully creating a doped array of single ions \cite{murdin2021error}. As the array size increases, the probability of failure in creating a perfect array of single ion implanted points also increases. However, it is feasible that devices may be constructed from imperfect arrays, where a number of sites have multiple ions or missing implantations and the use of such an imperfect array being countered by future quantum error correction codes. It is therefore useful to ensure as high a success rate as possible (without necessarily needing to be perfect) whilst also having some knowledge of regions where single, multiple, and no ions have been implanted. This will also provide a route to direct verification of the measured detection efficiency.

The ability to determine whether an ion has or has not been implanted, particularly on the single ion level, is material system dependent. For example, colour centres can be created in silicon \cite{hollenbach2022wafer} and diamond \cite{smith2019colour} by ion implantation. These systems, however, often have formation and activation efficiencies \cite{hollenbach2022wafer, pezzagna2010creation} and so are not completely representative of the results for direct implantation. 

Scanning tunnelling electron microscopy (STEM) can show atomic-scale incorporation of dopant ions within a thin membrane of material \cite{adshead2023high}. However the size of the areas under investigation are often too small (limited to 10s of nm at the most) for significant statistical data to be collated in experiments such as the ones conducted in this work. Novel STEM systems are being developed in order to scan larger regions of membranes with atomic resolution which will be invaluable in verification of the detection efficiency, however this will still be limited to membrane substrates with high-mass contrast between the membrane and implanted ion being necessary for high quality data. 

\subsection{Direct Peak Counting}

As $\lambda$ approaches 1 ion per pulse, the probability of having a single pulse containing more than one ion follows Poissonian statistics, which is to say that the probability of a pulse containing more than one ion increases significantly which is undesirable for the utilisation of the implanted ions as qubits, as unwanted ion-ion interaction can lead to qubit decoherence. This can be countered by reducing $\lambda$, for example, from 1 to 0.1 which reduces the probability of an ion pulse containing more than 1 ion from $\sim$26\% to $\sim$0.47\%. However, there are practical limitations to how small a $\lambda$ may be obtained, particularly regarding the source stability when operating in low beam current modes.

\begin{figure} [h!]
    \centering
 \subfloat[SED Peaks]
        {\includegraphics[width=0.5\textwidth]{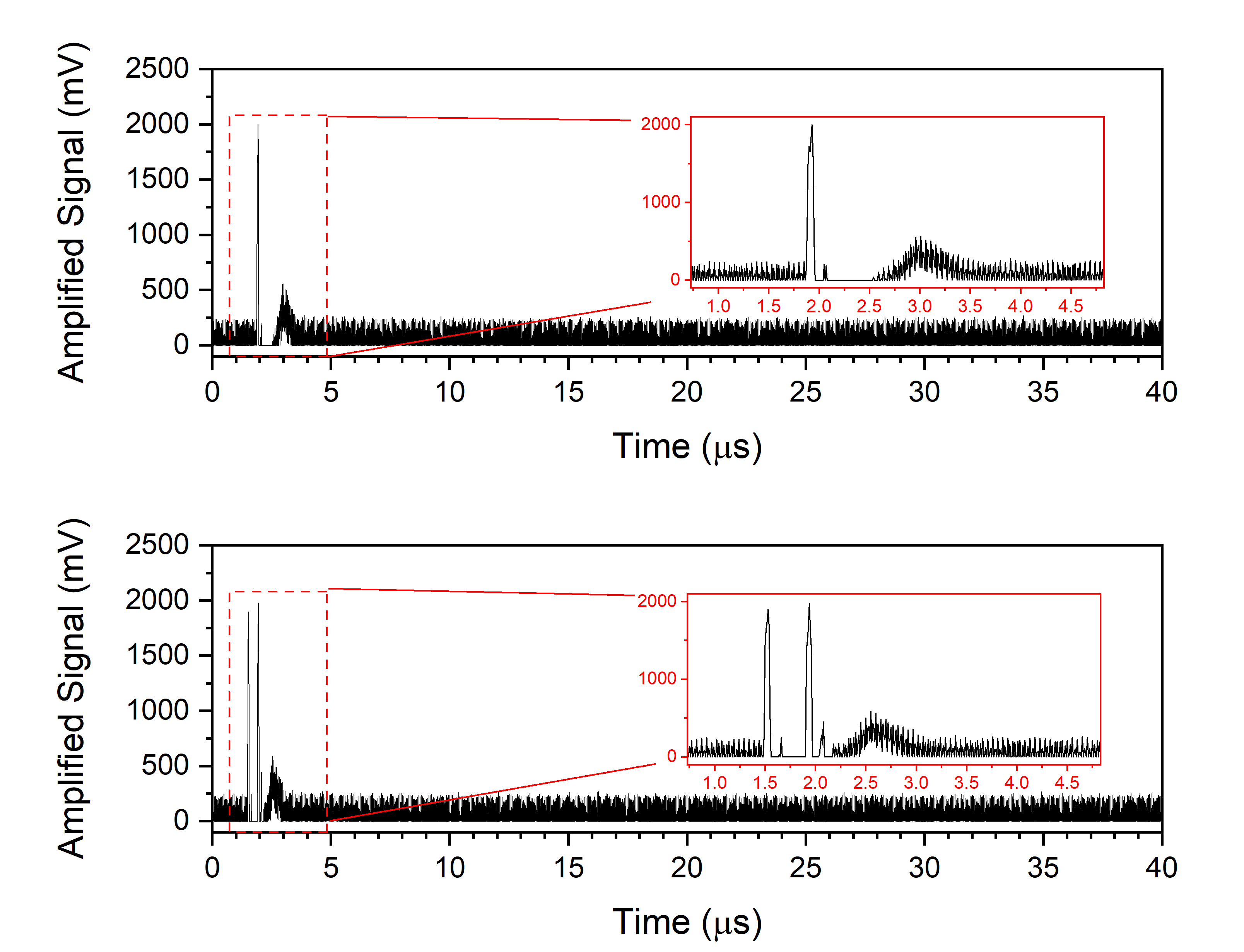}        \label{fig:peakCount:combo}}         \subfloat[Array] {\includegraphics[width=0.5\textwidth]{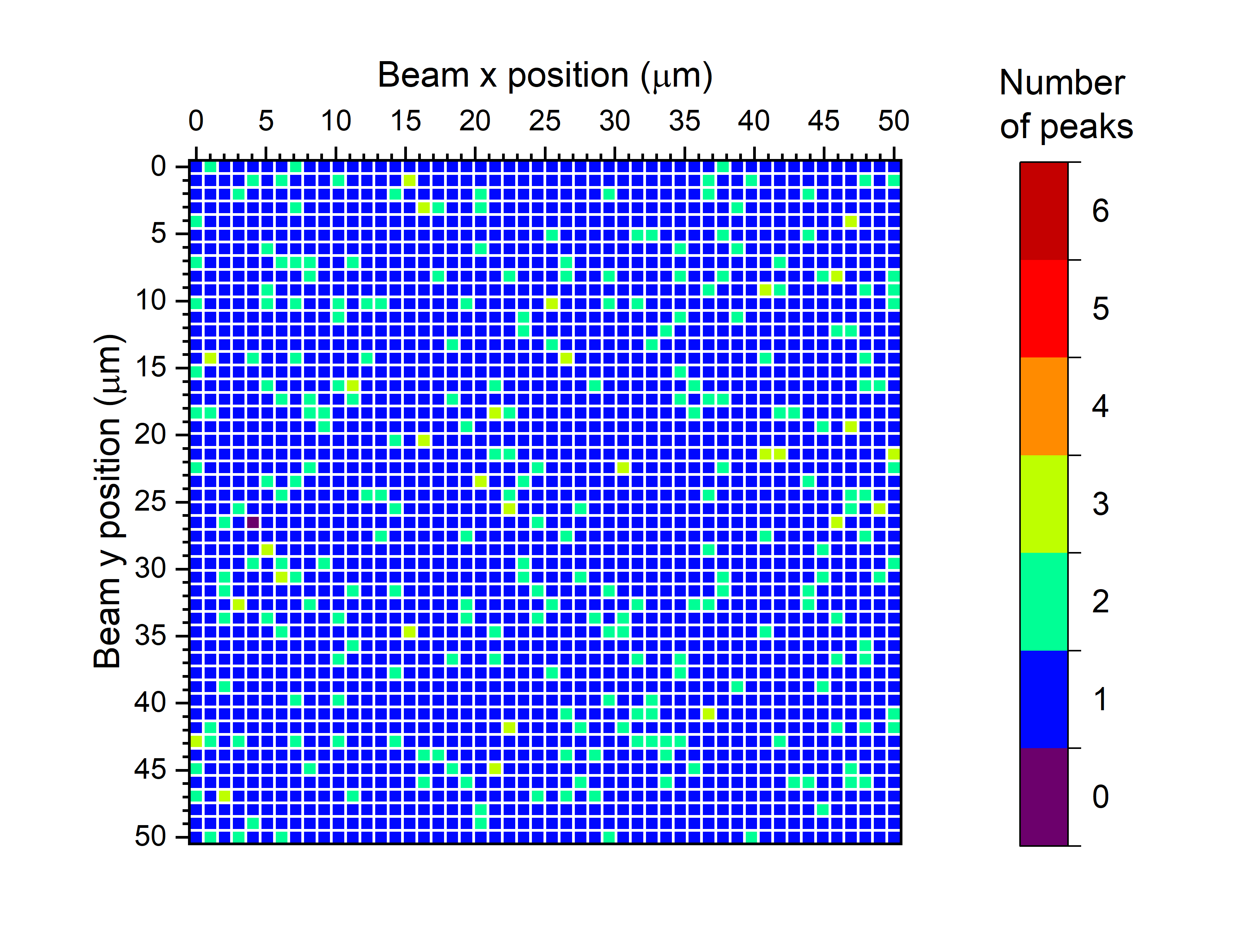}
        \label{fig:peakCount:array}} \\ 
    \caption{\subref{fig:peakCount:combo} Example SED traces showing single and double ion implantation events within a single pulse. The inset is a highlights the `detection window' which is the time within which a signal must be detected in order to register as a successful implantation event. \subref{fig:peakCount:array} Map of the number of ion implantation events detected at each point within a 50x50 point array during deterministic doping of 50 keV P into silicon.}
    \label{fig:peakCount}
\end{figure}

Even with the use of low values for $\lambda$, the ability to distinguish a single ion implantation from a multiple ion implantation event is desirable, which becomes possible with the utilisation of high-bandwidth SED electronics. Figure \ref{fig:peakCount:combo} provides examples of the detector output recorded for two ion pulses where successful detection occurred using the high-bandwidth detectors. One trace shows a single sharp peak corresponding to a single ion implantation, while the other shows two distinct sharp peaks, corresponding two ions being implanted during the single pulse. The insets in Figure \ref{fig:peakCount:combo} show the signals detected successfully within the `detection window'; the time within which a signal must be detected in order to register as a positive detection. This is to reduce the impact of dark counts (false positive detections) and to ensure the detections correlate with the pulsing the ion beam. The start time of the `detection window' scales with the momentum of the ion (the faster the ion, the sooner the `detection window' begins) and the width of the window correlates with the pulse length. 

Through counting the number peaks recorded within each detection window it is possible to map the number of implanted ions across the array for each value of $\lambda$. Figure \ref{fig:peakCount:array} shows such a map obtained for 50 keV P doping into silicon using $\lambda$ = 0.1 ions per pulse. Table \ref{tab:peakCount} shows that as $\lambda$ increases, the proportion of the 2500 successful detections which have a higher number of ion events increases, in line with what is expected from Poissonian statistics. 


\begin{table}[h!]
    \centering
    \begin{tabular}{c|c|c|c|c|c|c|c}
    $\lambda$ & \multicolumn{7}{c}{\% of pulses with $j$ peaks, experimental (Poissonian)}  \\
    (ions / pulse) & 0 & 1 & 2 & 3 & 4 & 5 & 6  \\
    \hline
    0.0953 & 93 (91) & 6 (9) & 1 (0) & 0 (0) & 0 (0) & 0 (0) & 0 (0) \\
    0.1923 & 86 (83) & 12 (16) & 2 (2) & 0 (0) & 0 (0) & 0 (0) & 0 (0) \\
    0.2893 & 78 (75) & 18 (22) & 3 (3) & 1 (0) & 0 (0) & 0 (0) & 0 (0) \\
    0.3863 & 71 (68) & 23 (26) & 5 (5) & 1 (1) & 0 (0) & 0 (0) & 0 (0) \\
    0.4833 & 65 (62) & 26 (30) & 7 (7) & 1 (1) & 0 (0) & 0 (0) & 0 (0) \\
    0.5803 & 60 (56) & 29 (32) & 9 (9) & 2 (2) & 1 (0) & 0 (0) & 0 (0) \\
    0.6773 & 55 (51) & 31 (34) & 11 (12) & 3 (3) & 1 (0) & 0 (0) & 0 (0) \\
    0.7743 & 51 (46) & 32 (36) & 12 (14) & 3 (4) & 1 (1) & 0 (0) & 0 (0) \\
    0.8713 & 44 (42) & 35 (36) & 14 (16) & 5 (5) & 1 (1) & 0 (0) & 0 (0) \\
    0.9683 & 42 (38) & 33 (37) & 16 (18) & 6 (6) & 2 (1) & 0 (0) & 0 (0) \\
    \end{tabular}
    \caption{Percentage of total pulses for each value of $\lambda$ which contain a particular number of peaks for the 50 keV P implantation into silicon dataset. Numbers in the brackets are the percentage probabilities of a pulse containing that  particular number of ions, given by Poissonian statistics.}
    \label{tab:peakCount}
\end{table}


There is a discrepancy between the experimentally determined percentage of pulses containing a particular number of peaks and that predicted by Poissonian statistics. It is possible, however, to adjust the predicted (Poissonian) values with a scalar ($S$) such that a closer agreement with the experimentally observed values is found. Equation \ref{eq:peakScale:nonZero} shows the scaled Poisson probability as a function of number of ions, $p(j)$, per pulse that is non-zero. Equation \ref{eq:peakScale:zero} takes into account the proportion of predicted pulses which are missed by the detector, and therefore registered as empty pulses.

\begin{subequations}
    \begin{equation}
        p(j)=S\left(\frac{\lambda^j\text{exp}(-\lambda)}{j!}\right), \hspace{50pt} j>0
        \label{eq:peakScale:nonZero}
    \end{equation}
    \begin{equation}
        p(0)=\text{exp}({-\lambda})(1-S)\sum_{j=1}^6\left(\frac{\lambda^j\text{exp}(-\lambda)}{j!}\right)
        \label{eq:peakScale:zero}
    \end{equation}
    \label{eq:peakScale}
\end{subequations} 

The scalar for which the difference between the experimental values and the scaled Poisonian values is smallest (maximum difference of 2.34\%, Tables S.2 
and S.3
in Supporting Information) is 0.93, indicating a 93\% confidence in detection and peak prediction. This method of determining the number of ions implanted into a substrate assumes that each ion resulting from a pulse arrives at the substrate, independent of the rest of the ions in a given pulse and does so within the set pulse width (typically greater than 200 ns, as per Table \ref{tab:exampleDetection}). This approach requires further investigation in order to be confirmed as a robust method of direct peak counting, including independent verification of the number of ions implanted into each spot each spot which may be achieved using a suitable method such as scanning transmission electron microscopy.

%% file: Sections/ResultsAndDiscussion.tex
\section{Detection Efficiency}

The detection efficiencies were determined for a variety of ions, implanted into a number of different substrates, using the linear regression method outlined above. For each ion species two ion acceleration voltages of 12.5 and 25 kV were used, with the latter being the maximum anode voltage of the P-NAME tool used for this work \cite{adshead2023high}. Selection of the ion species' charge state using the Wien filter enabled access to ion implantation energies of 12.5 keV (singly-charged ion, 12.5 kV anode), 25 keV (either doubly-charged ion and 12.5 kV anode, or singly-charged ion and 25 kV anode) and 50 keV (doubly-charged ion, 25 kV anode). Ion cluster implantation (e.g. Au$^+_2$) was also investigated for a limited number of cases to further explore the relationship between ion beam configuration and detection efficiency.

\subsection{Silicon} 

\begin{figure} [h!]
    \centering
    \subfloat[Energy comparison]
        {\includegraphics[width=0.5\textwidth]{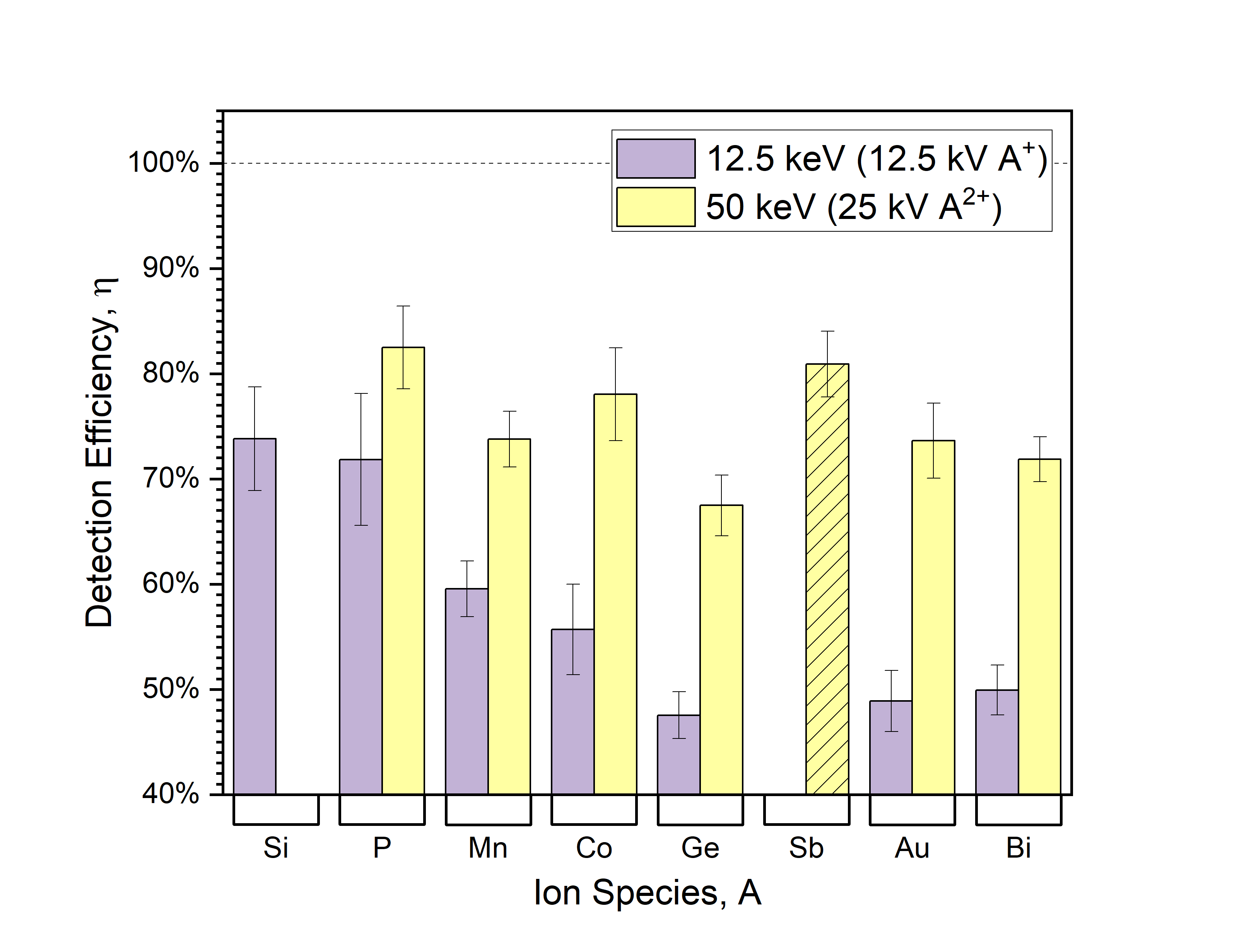}
        \label{fig:si:energyVar}}\subfloat[Charge state comparison]
        {\includegraphics[width=0.5\textwidth]{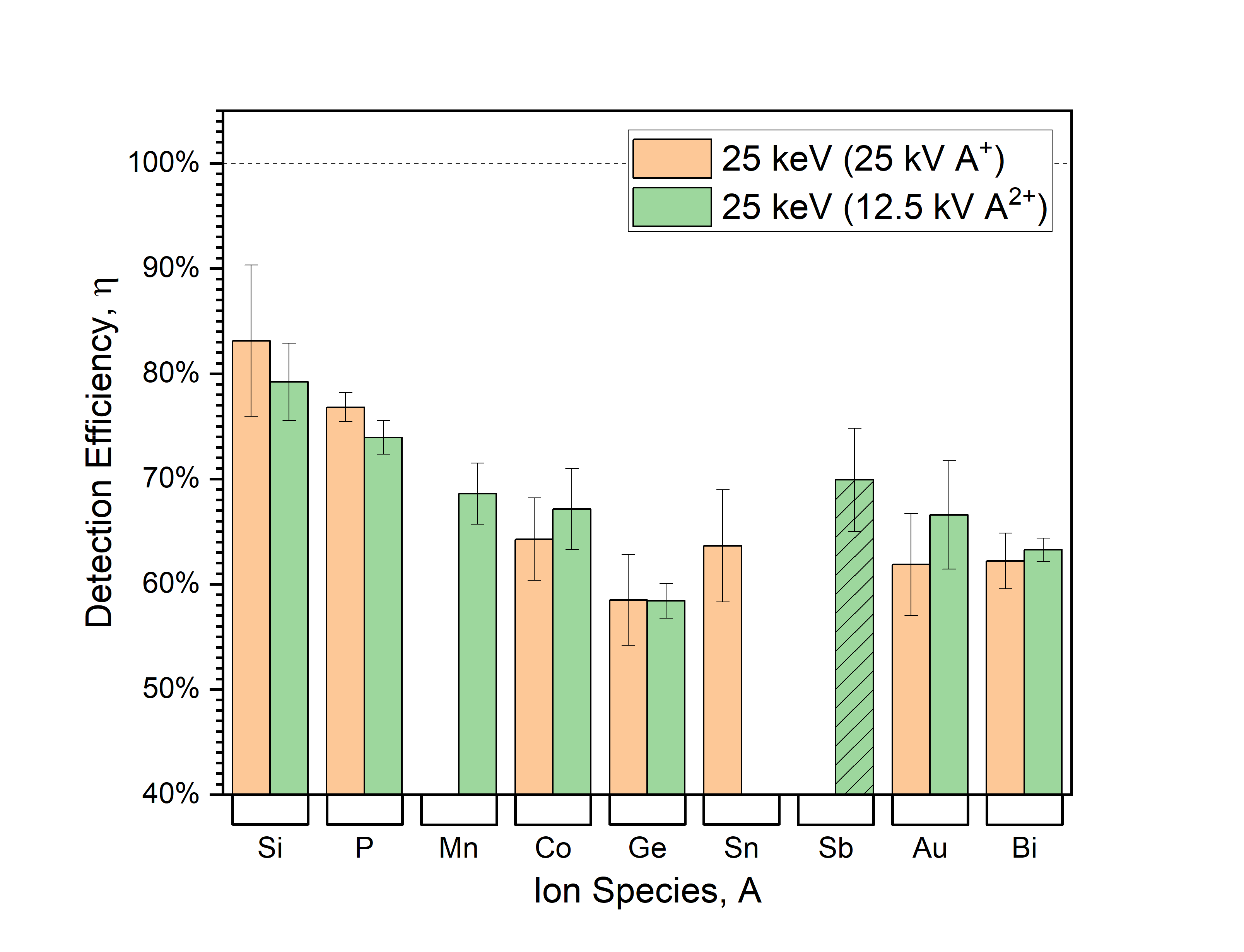}        \label{fig:si:chargeVar}}
    \caption{Detection efficiency measurements for various ion species into undoped silicon (through a native oxide layer). The anode voltage and charge state of the ions was varied to obtain different implantation energies \subref{fig:si:energyVar} and the same implantation energy \subref{fig:si:chargeVar}. Hashed bars represent data that was previously reported \cite{adshead2025isotopically}.}
    \label{fig:si_detections}
\end{figure}

Figure \ref{fig:si:energyVar} shows the measured detection efficiencies for the various ion species when implanted into silicon (through a native oxide layer). Increasing the implantation energy from 12.5 keV to 50 keV results in an increase in the detection efficiency across all the ion species investigated. 

In Figure \ref{fig:si:chargeVar}, a common implantation energy of 25 keV is achieved using two different anode voltages by selecting the charge of the ions. While there is a variation in the absolute value of detection efficiency between the singly- and doubly-charged ion states, in the cases for which a comparison may be made (Si, P, Co, Ge, Au and Bi), the figures fall within each others error bounds.

\subsection{Silicon Dioxide} 

\begin{figure} [h!]
    \centering
    \subfloat[Energy comparison]
        {\includegraphics[width=0.5\textwidth]{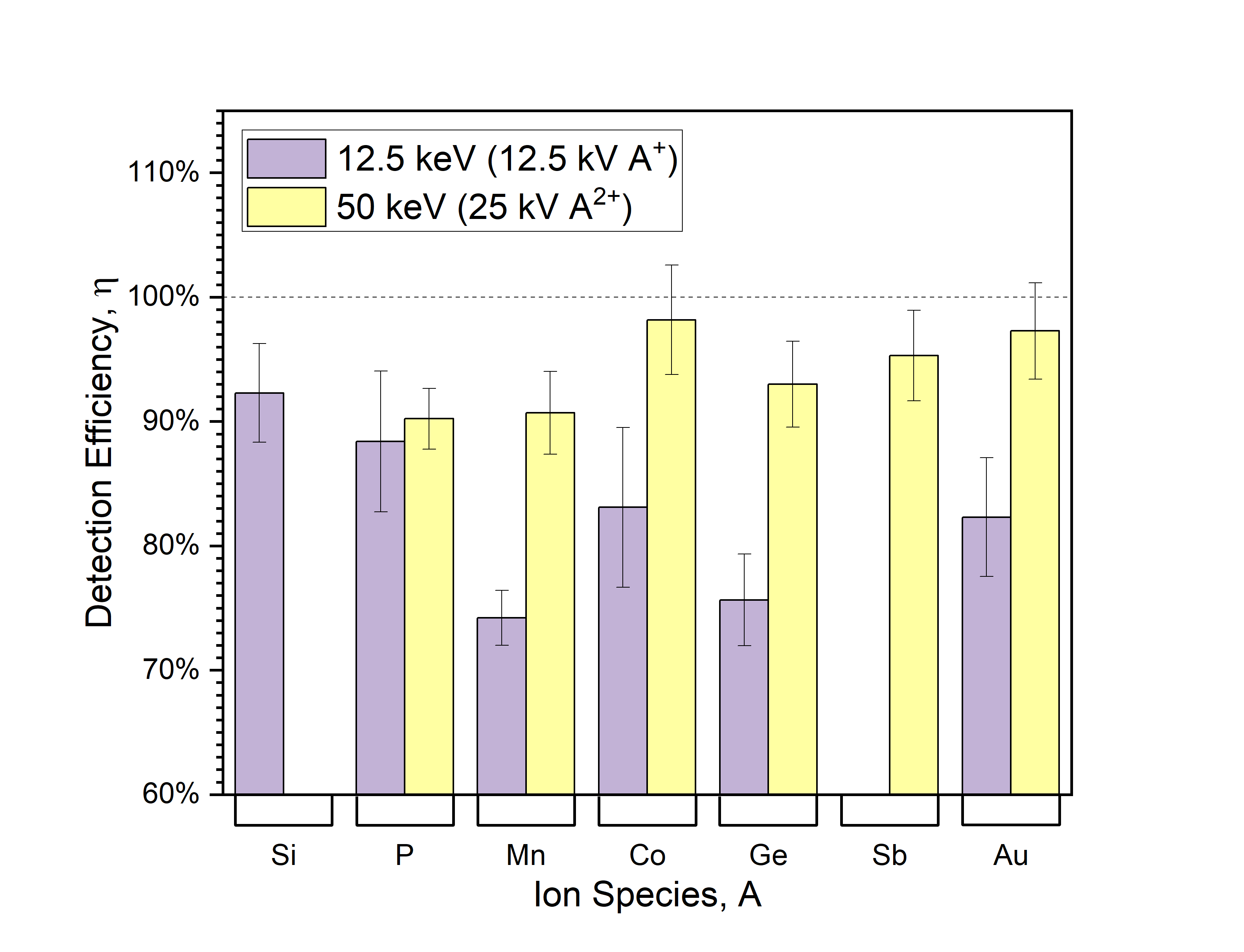}
        \label{fig:sio2:energyVar}}\subfloat[Charge state comparison]
        {\includegraphics[width=0.5\textwidth]{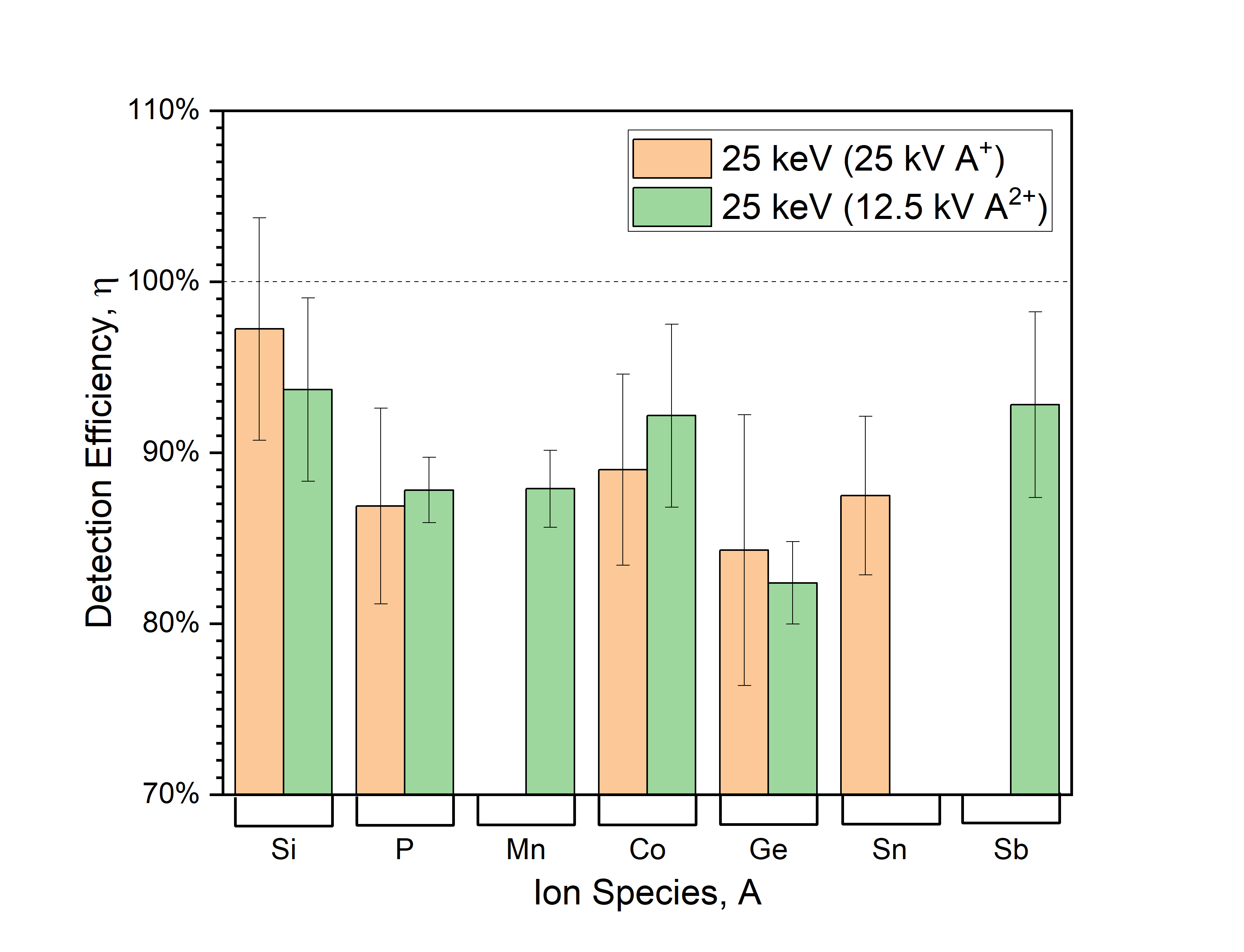}        \label{fig:sio2:chargeVar}}
    \caption{Detection efficiency measurements for various ion species into SiO$_2$. The anode voltage and charge state of the ions was varied to obtain different implantation energies \subref{fig:sio2:energyVar} and the same implantation energy \subref{fig:sio2:chargeVar}.}
    \label{fig:sio2_detections}
\end{figure}

Figure \ref{fig:sio2_detections} shows the measured detection efficiencies for the various ion species implanted into SiO\textsubscript{2}. The detection efficiency values all fall within the range of 70-100 \% for all the ion beam species explored. As the implantation energy is increased, there is an increase in detection efficiency (Figure \ref{fig:sio2:energyVar}). The effect of ion charge state variation, for the same implantation energy, (Figure \ref{fig:sio2:chargeVar}) reveals no distinct variation in detection efficiency for the two charge states beyond the error bounds of the measurements. 

Comparing the measured detection efficiencies between silicon and SiO\textsubscript{2} (200 nm SiO\textsubscript{2} on boron-doped silicon, see Experimental Methods for more details) reveals that all of the efficiencies in the oxide sample are greater than those measured in silicon with only a native oxide. Since the oxide thickness is too great for any of the ions of the energies investigated in this work to pass through the oxide layer and stop in the underlying silicon layer of the SiO\textsubscript{2}, the increase in detection efficiency is expected to result from the oxide nature of the substrate.

\subsection{Aluminium Oxide} 

\begin{figure} [h!]
    \centering
    \subfloat[Energy comparison]
        {\includegraphics[width=0.5\textwidth]{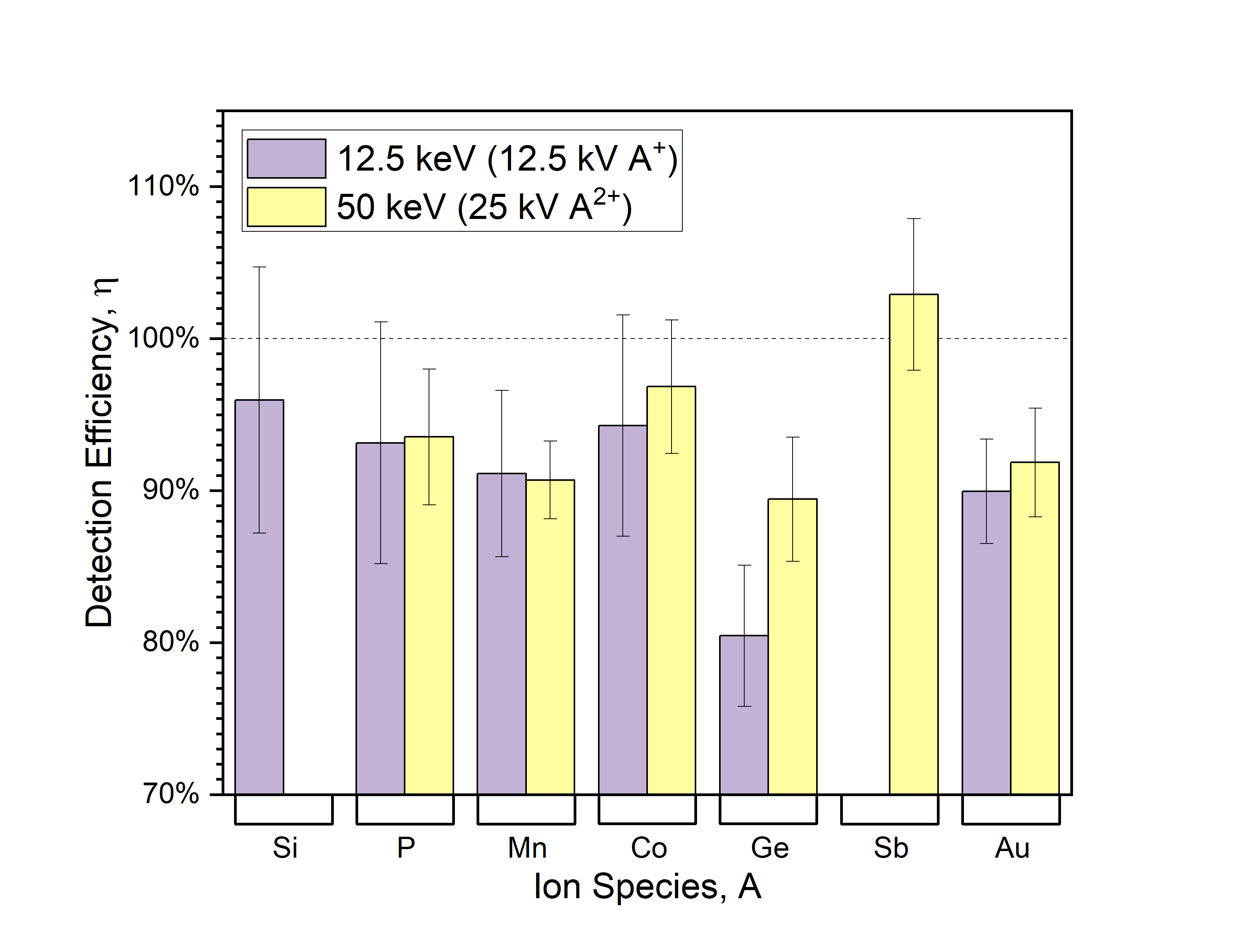}
        \label{fig:al2o3:energyVar}}\subfloat[Charge state comparison]
        {\includegraphics[width=0.5\textwidth]{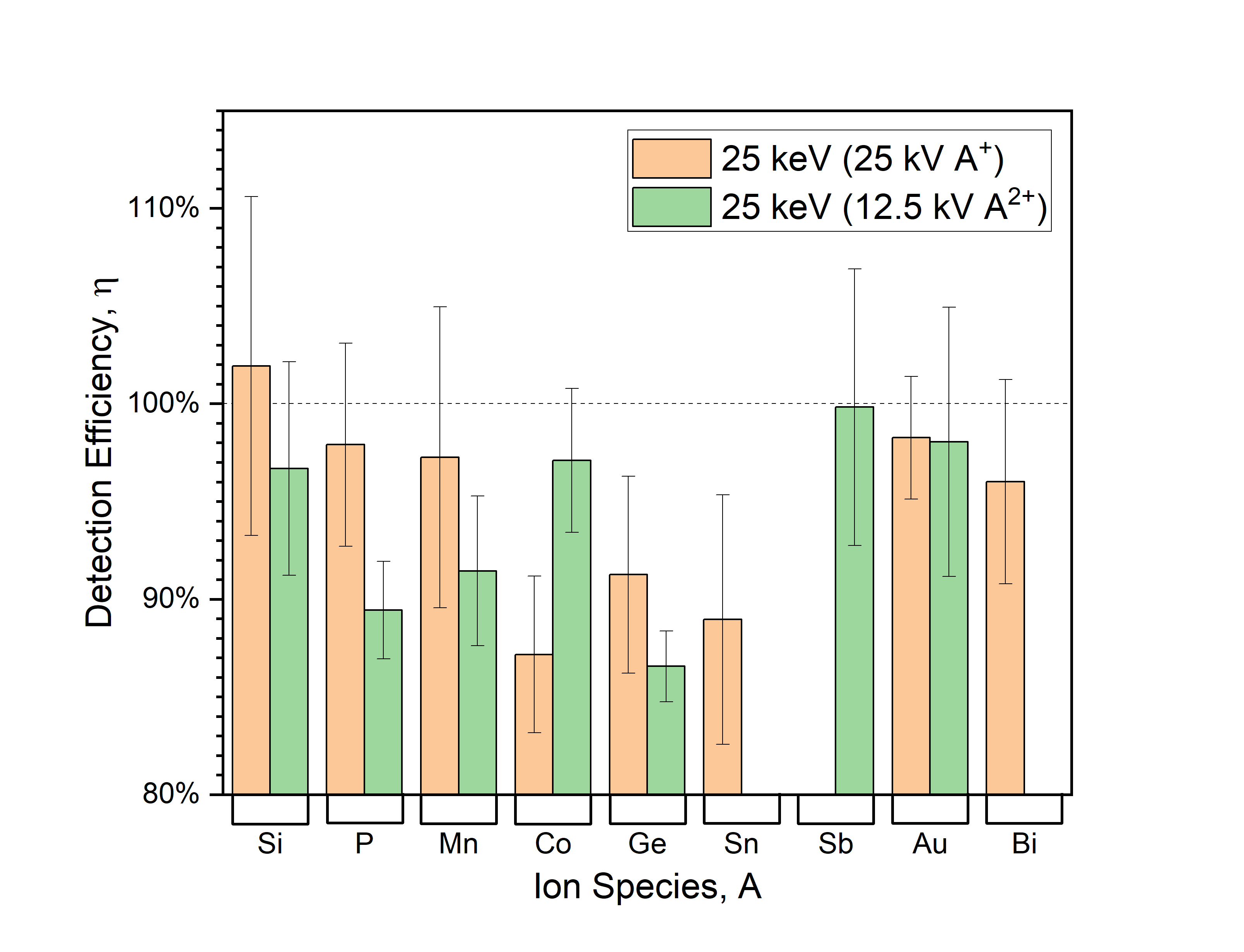}        \label{fig:al2o3:chargeVar}}
    \caption{Detection efficiency measurements for various ion species implanted into Al\textsubscript{2}O\textsubscript{3}. The anode voltage and charge state of the ions was varied to obtain different implantation energies \subref{fig:al2o3:energyVar} and the same implantation energy \subref{fig:al2o3:chargeVar}.}
    \label{fig:al2o3_detections}
\end{figure}

In the case of Al\textsubscript{2}O\textsubscript{3}, Figure \ref{fig:al2o3:energyVar} reveals that for a number of ions there is not a significant increase in detection efficiency as the energy is increased, the only exception being Ge. Some determined values of the detection efficiency (e.g. 50 keV Sb and  25 keV Si, singly-charged variation) are above 100\%, though the lower error bounds bring the value below 100\%.

The effect of changing ion charge state  (Figure \ref{fig:al2o3:chargeVar}) shows greater variation in the detection efficiencies, with the majority of singly-charged ions (Si, P, Mn, and Ge) all having higher detection efficiencies than their doubly-charged counterparts. Conversely, Co stands out as the sole species with a higher detection efficiency when a doubly-charged ion is implanted. Au in Al\textsubscript{2}O\textsubscript{3}, as with the silicon substrate, has very similar absolute detection efficiency values, however for the doubly-charged ion the error-bounds are more significant.

The detection efficiencies measured for implantation into Al\textsubscript{2}O\textsubscript{3} are comparable to, if not greater than, those measured for SiO\textsubscript{2}. This indicates that the oxide nature of these two substrates results in a higher secondary electron yield and therefore greater detection efficiency for the ions investigated here, relative to the non-oxide substrates.

\subsection{Gallium Arsenide} 

\begin{figure} [h!]
    \centering
    \subfloat[Energy comparison]
        {\includegraphics[width=0.5\textwidth]{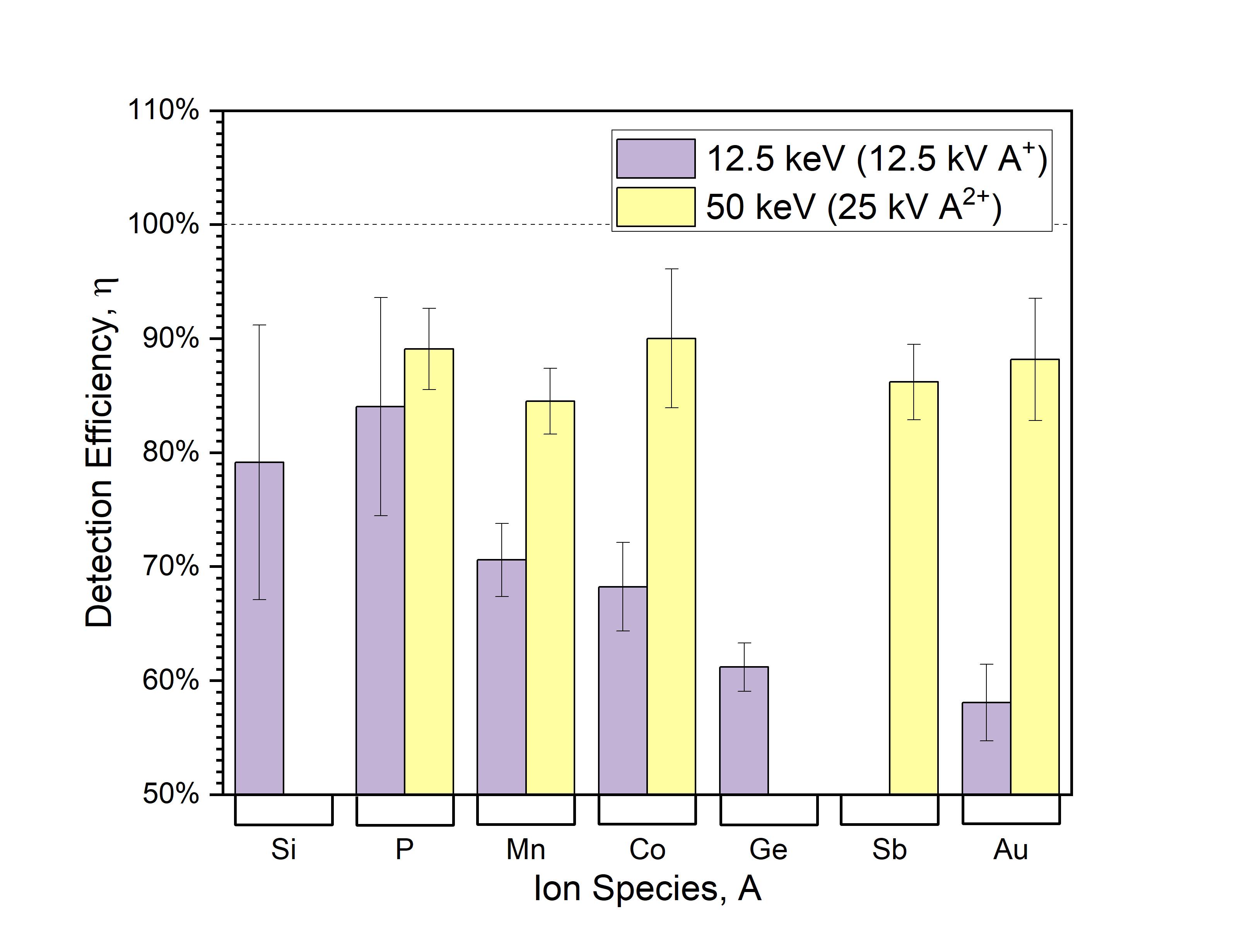}
        \label{fig:gaas:energyVar}}\subfloat[Charge state comparison]
        {\includegraphics[width=0.5\textwidth]{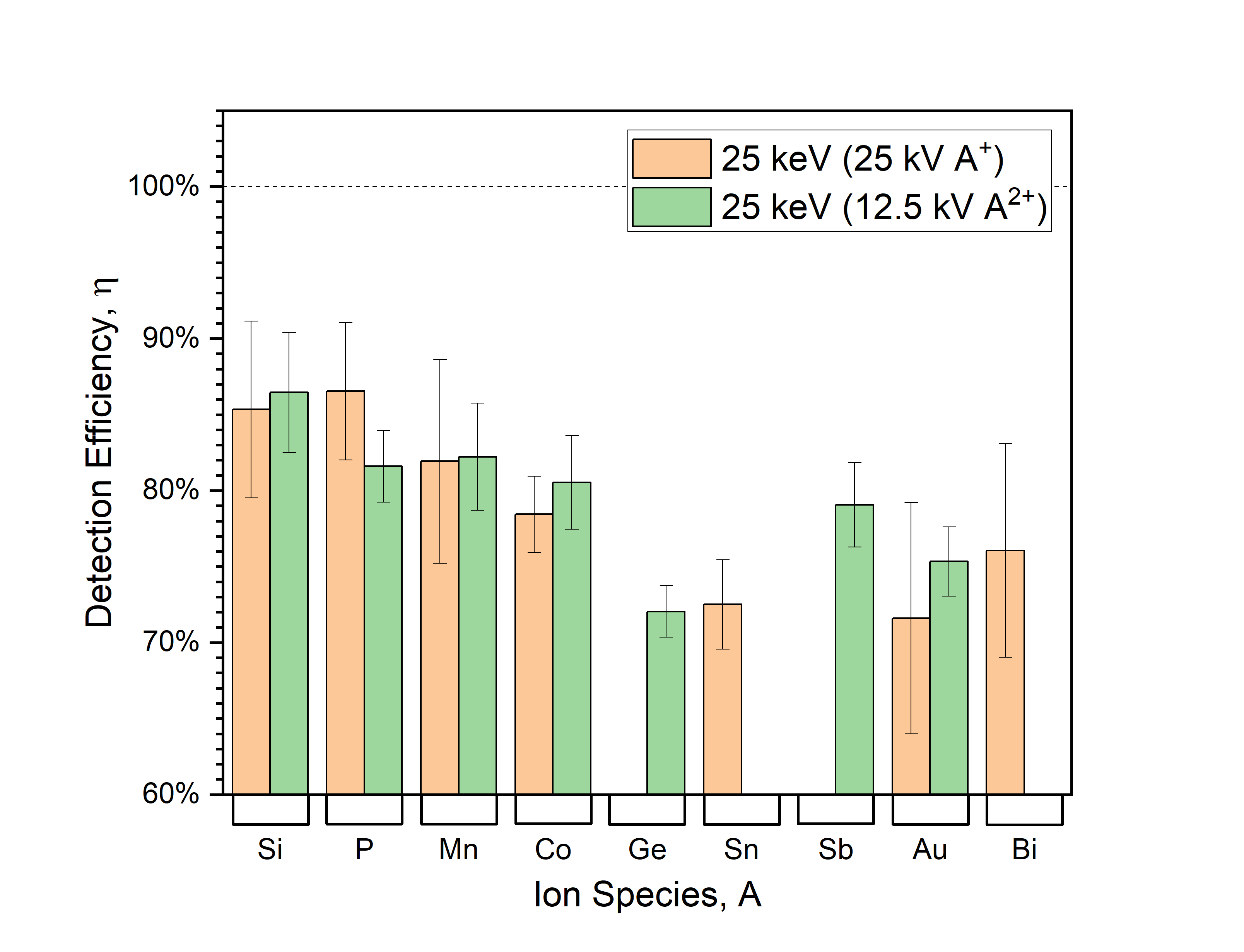}        \label{fig:gaas:chargeVar}}
    \caption{Detection efficiency measurements for various ion species implanted into GaAs. The anode voltage and charge state of the ions was varied to obtain different implantation energies \subref{fig:gaas:energyVar} and the same implantation energy \subref{fig:gaas:chargeVar}.}
    \label{fig:gaas_detections}
\end{figure}

There is a significant increase in the measured detection efficiency for ion implantation into GaAs as the implantation energy increases. Furthermore, the magnitude of this increase is greater as the ion mass increases (Figure \ref{fig:gaas:energyVar}), with Au showing the greatest increase. 

Considering the error bounds in Figure \ref{fig:gaas:chargeVar}, the variation of ion charge state does not have a significant impact on the detection efficiency, with the possible exception of P, which has a lower efficiency for the doubly-charged case.

\subsection{Diamond}

\begin{figure} [h!]
    \centering
    \subfloat[Energy comparison]
        {\includegraphics[width=0.5\textwidth]{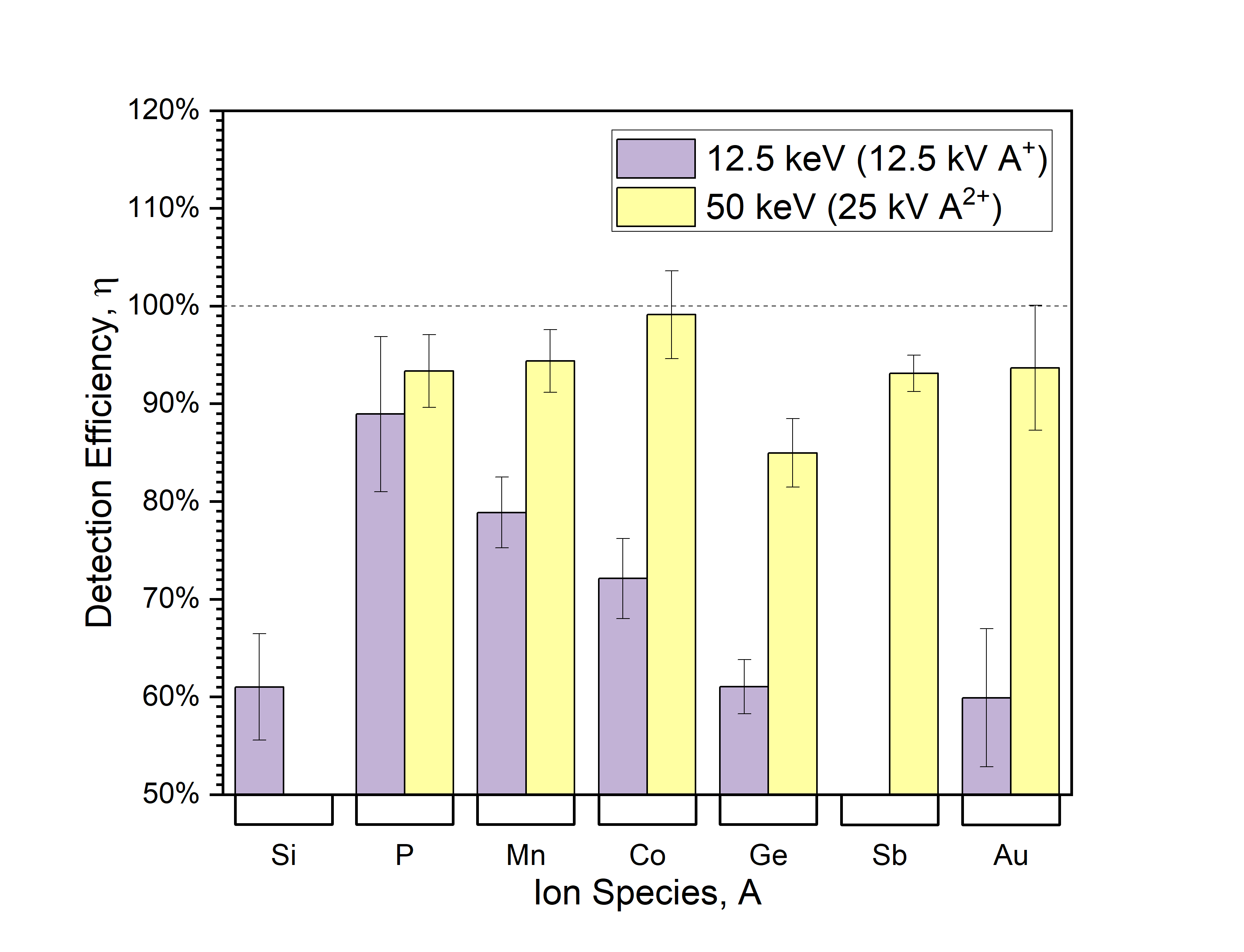}        \label{fig:diamond:energyVar}}\subfloat[Charge state comparison]
        {\includegraphics[width=0.5\textwidth]{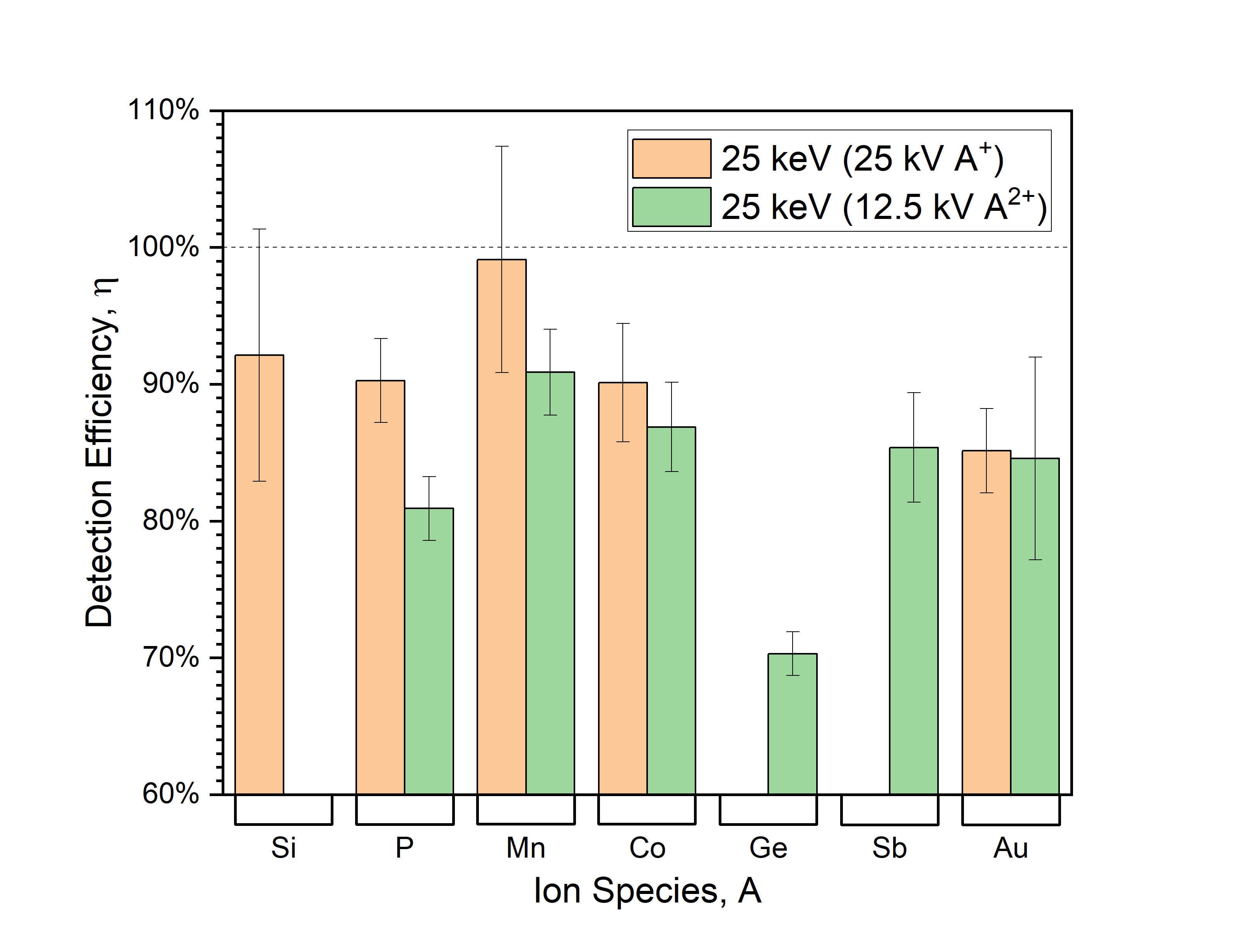}        \label{fig:diamond:chargeVar}}
    \caption{Detection efficiency measurements for various ion species implanted into diamond. The anode voltage and charge state of the ions was varied to obtain different implantation energies \subref{fig:diamond:energyVar} and the same implantation energy \subref{fig:diamond:chargeVar}.}
    \label{fig:diamond_detections}
\end{figure}

Figure \ref{fig:diamond:energyVar} shows that P implantation into diamond does not demonstrate a significant increase in detection efficiency as the energy is increased, while the rest of the ions investigated do. However, P shows a statistically significant decrease in detection efficiency when the charge state is increased for the same implantation energy (Figure \ref{fig:diamond:chargeVar}). Mn, Co and Au also show a decrease in efficiency as the charge state is increased, though within their respective error bounds.

Of the group IV elements, that are useful for creating colour centres in diamond, only Si and Ge have results presented here, with 25 keV Si (achieved with singly-charged 25 kV Si) showing the greatest detection efficiency ($\sim$92\%), though with a large error. With such a high detection efficiency, it is of interest to investigate deterministic implantation with large-scale arrays with methods of deterministic colour centre activation, particularly laser annealing \cite{cheng2025laser}. Combining deterministic ion implantation with with direct peak counting and direct laser annealing will enable a full exploration of the defect complexes which form with a specific number of ions being incorporated into the diamond substrate with nanoscale precision achievable with FIB systems.

\subsection{Silicon Carbide} 

\begin{figure} [h!]
    \centering
    \subfloat[Energy comparison]
        {\includegraphics[width=0.5\textwidth]{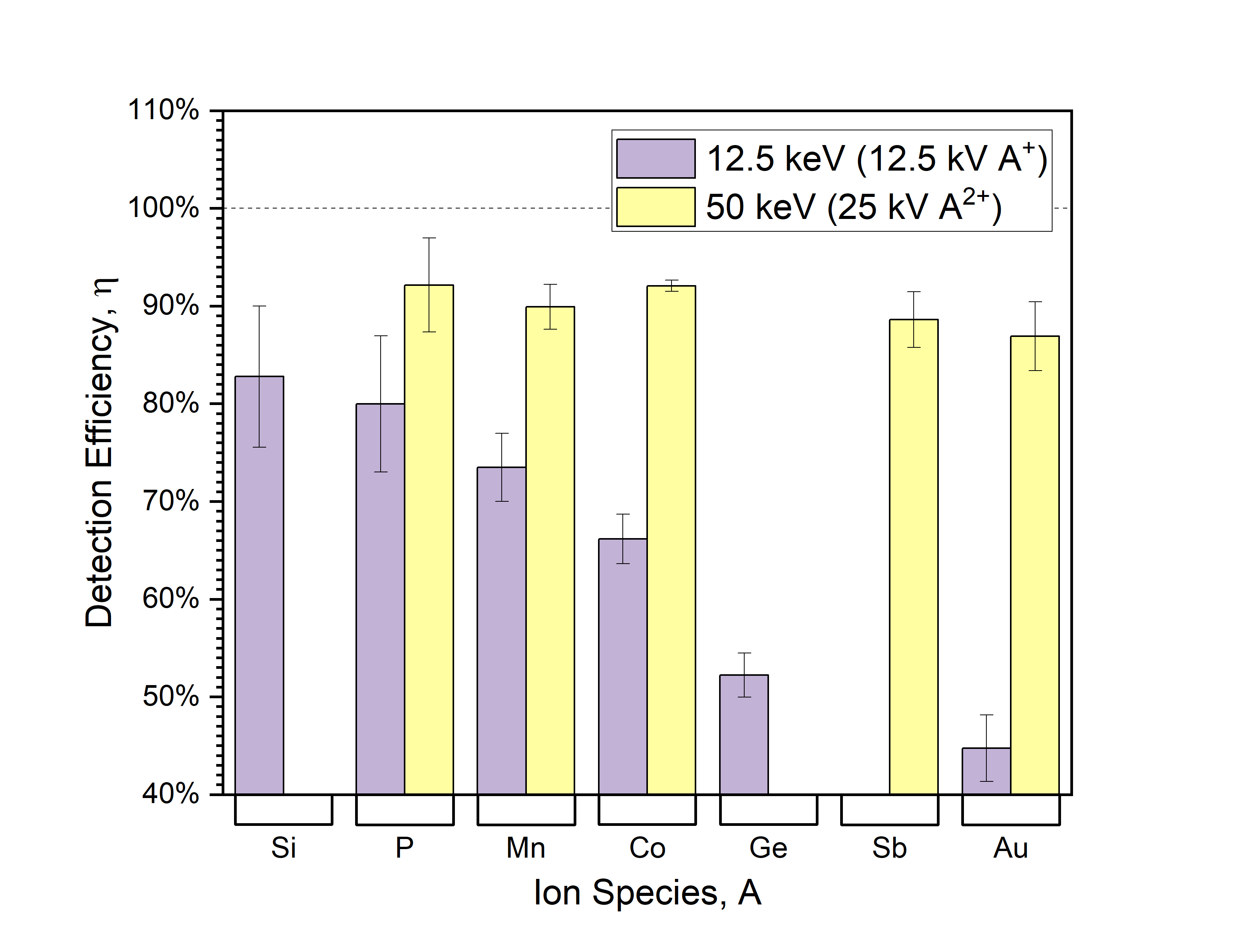}        \label{fig:sic:energyVar}}\subfloat[Charge state comparison]
        {\includegraphics[width=0.5\textwidth]{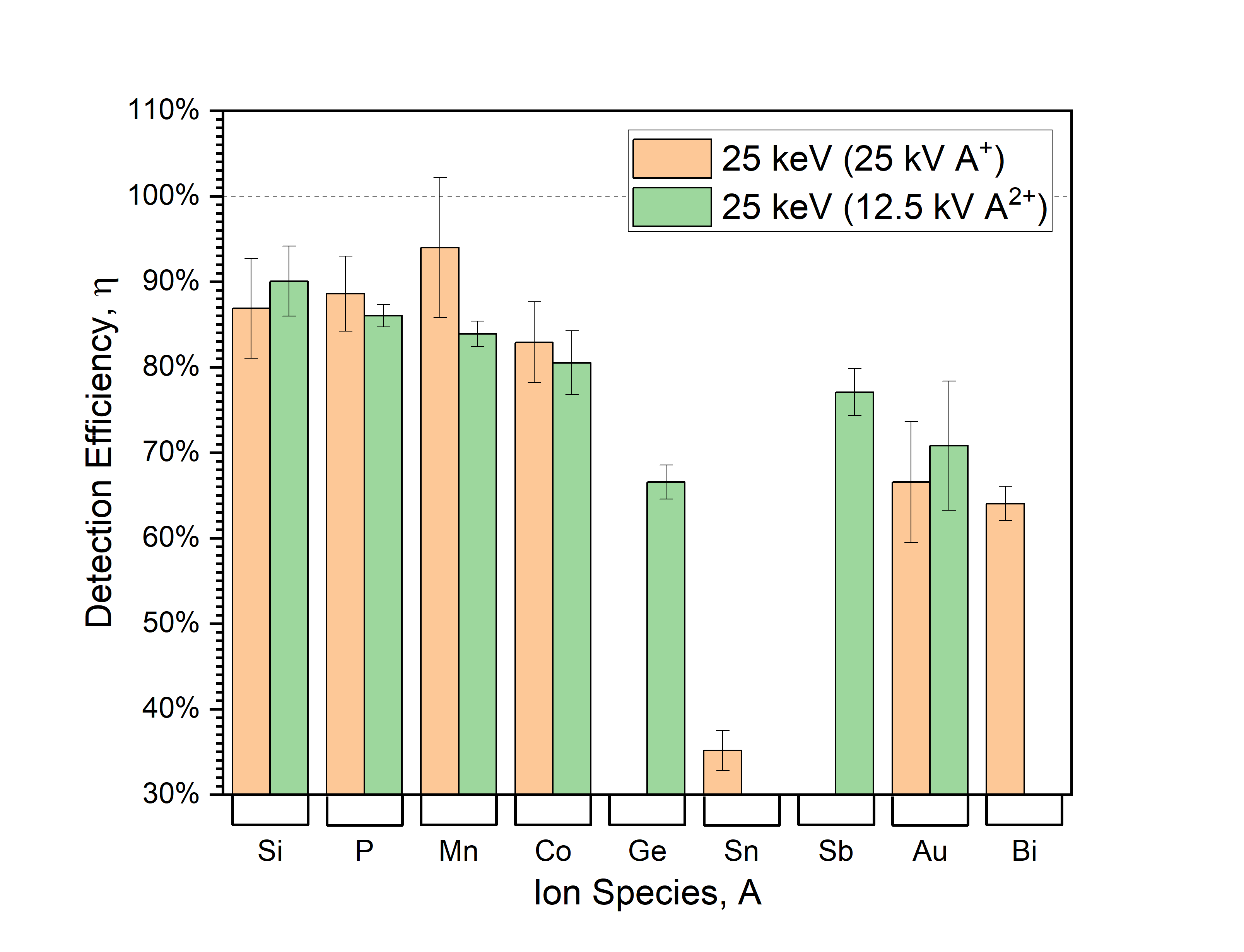}\label{fig:sic:chargeVar}}
    \caption{Detection efficiency measurements for various ion species impllanted into silicon carbide. The anode voltage and charge state of the ions was varied to obtain different implantation energies \subref{fig:sic:energyVar} and the same implantation energy \subref{fig:sic:chargeVar}.}
    \label{fig:sic_detections}
\end{figure}

The measured detection efficiencies of single ion implantation into SiC are seen to decrease with increasing ion mass when comparing fixed implantation energies, Figure \ref{fig:sic:energyVar}. This trend is seen for implantation into other substrates but not as clearly as for SiC. Increasing the implantation energy is consistently found to improve the detection efficiency for all ions implanted into SiC, with the greatest difference being observed between the 12.5 keV and 50 keV Au implantation. Figure \ref{fig:sic:chargeVar} shows that, with the exception of Mn, there is no significant difference in the detection efficiencies of singly- and doubly-charged ions. 

SiC, like diamond, can form colour centres attributed to defects and vacancies which may be introduced through ion implantation \cite{kobayashi2021intrinsic}. Similar to diamond, the ability to implant single ions with high detection efficiencies can provide the opportunity to investigate single and ensemble defects in SiC, and the emission characteristics thereof for engineering emitters for qubit applications at room temperature \cite{widmann2015coherent}.

\subsection{Cluster Implantations}

\begin{figure} [h!]
    \centering
    \includegraphics[width=0.8\textwidth]{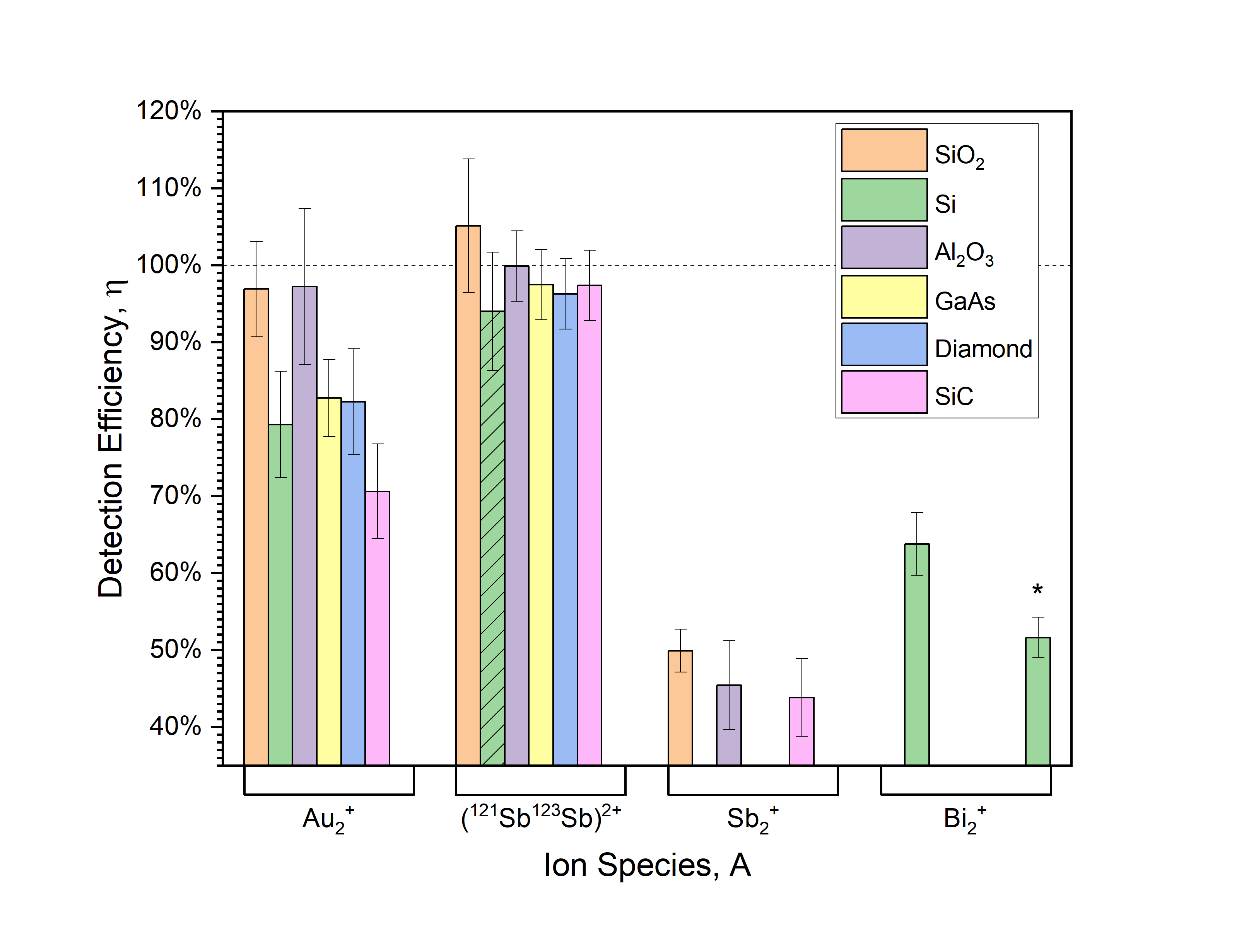} 
    \caption{Detection efficiency measurements for various ion species, A, into different substrates defined in the legend. Hashed bars represent data that was previously reported \cite{adshead2025isotopically}. All ion clusters were implanted with an anode of 25 kV, with the exception of the second Bi$_2^+$ (demarcated with an asterisk, *), which was implanted with an anode voltage of 12.5 kV.}
    \label{fig:cluster_detections}
\end{figure}

The detection efficiency for ion cluster implantation was investigated non-exhaustively as part of this work. For all of the ion clusters studied an anode voltage of 25 kV was chosen, with the exception of Bi$_2^+$, which was also investigated for 12.5 kV implantation into Si (demarcated with an asterisk in Figure \ref{fig:cluster_detections}).

There are two Sb clusters presented in Figure \ref{fig:cluster_detections}, one of which is a typical cluster formed from two of the same isotope (specifically \textsuperscript{123}Sb, which was the Sb isotope used in the single-ion detection efficiency measurements in this work). The other, however, is a unique doubly-charged cluster formed from a single ion of each Sb isotope (\textsuperscript{121}Sb\textsuperscript{123}Sb), which has been previously reported \cite{adshead2025isotopically}. The hashed bar in Figure \ref{fig:cluster_detections}, and in the silicon data presented above in Figure \ref{fig:si_detections}, represents detection efficiency data we have previously reported.

Comparison of the ion cluster detection efficiency data presented in Figure \ref{fig:cluster_detections} with the single-ion data presented within is possible by considering the energy per atom for the species being implanted. For example, a 25 keV Au$_2^+$ is presumed to implant with 12.5 keV per atom within the cluster, which can be compared with the 12.5 keV Au\textsuperscript{+} ion implantation presented above. When using this comparison, the 25 kV clusters Au$_2^+$, (\textsuperscript{121}Sb\textsuperscript{123}Sb)\textsuperscript{2+}, and Bi$_2^+$ (non-asterisk data in Figure \ref{fig:cluster_detections}) can be compared with 12.5 kV ions Au\textsuperscript{+}, Sb\textsuperscript{2+}, and Bi\textsuperscript{+}, respectively.

Doing so it is observed that most of the detection efficiencies for the ion-clusters are greater than for their single-ion counterpart. The greatest variation was found in the case of 25 kV Au$_2^+$ compared to 12.5 kV Au\textsuperscript{+} (both 12.5 keV per atom) into silicon, where the cluster was found to lead to a 30\% increase in detection efficiency (79 ($\pm$7)\% and 49 ($\pm$3)\%, respectively).

The smallest (non-zero) difference is found for 25 kV Au$_2^+$ implantation into in Al\textsubscript{2}O\textsubscript{3}, where  a 7\% increase in detection efficiency was found when comparing to the 12.5 kV Au\textsuperscript{+}, with efficiencies of 97 ($\pm$10)\% and 90 ($\pm$3)\%, respectively. No difference was found between 25 kV (\textsuperscript{121}Sb\textsuperscript{123}Sb)\textsuperscript{2+} compared to 12.5 kV Sb\textsuperscript{2+} in Al\textsubscript{2}O\textsubscript{3}, where both resulted in a value of 100 ($\pm$5)\%.

Whilst there is an increase in detection efficiency for cluster implantations relative to the single-ion implantations, for applications in single atom devices, like qubits following the Kane proposal, implanting clusters will guarantee that the ion of interest for encoding of the qubit state will be in close proximity to a second ion, resulting in decoherence. This is assuming that the two ions within the cluster separate by the straggle predicted by the energy per atom in the cluster implantation, which is typically on the order of a few nm in the case of the energies explored in this work \cite{adshead2023high, adshead2025isotopically}.

In cases of cluster ion implantation, the high efficiency deterministic implantation of ions with separation on the order of nm can be advantageous for qubit devices, with proximal implantations allowing for overlapping electron wavefunctions. This allows for the coupling of the nuclear spins of the adjacent ions via the shared electron, resulting in a three-qubit system, previously demonstrated with pairs of P implanted into silicon \cite{mkadzik2022precision}. While P clusters have not been considered in the work presented here, the P-NAME system is able to access such a cluster and is expected to follow the trend of increased detection efficiency for the cluster compared to the single-ion implantation (72 ($\pm$6) \% for 12.5 keV P in Si).

Another ion species of interest for solid-state qubit systems, particularly in Si, is Sb. The potential for application of mono-atomic and coupled (by cluster implantation) Sb qubits in Si has been discussed previously \cite{adshead2025isotopically}, with the coupling of the high nuclear spin Sb ions via a shared electron expected to realise a larger Hilbert space onto which quantum information may be encoded than the coupled P or mono-atomic Sb systems.

\section{Discussion}

The most important factor influencing the detection efficiency is the secondary electron yield, which is related to the implanted ion species and substrate combination. However, as the SEDs are comprised of channeltron electron multipliers in order to amplify the secondary electrons to an appreciable (and detectable) signal, information about the secondary electron emission (number of secondary electrons, their energy and direction of emission, etc.) is not available. However, comparisons between the detection efficiencies obtained here and previous trends in secondary electron emission may still be made. 

The trend of increasing secondary electron yield as implantation energy \cite{chutopa2003measurement} and cluster size \cite{baudin1996sublinear} are increased is well established for a range of ion species and substrate combinations. Previous studies into secondary electron emission with various gaseous ion species and charge states have seen an increase in secondary electron emission as the charge state is increased \cite{eder1997precise}. In contrast, the charge state variation investigated in this study, using non-gaseous species, does not result in a significant increase in detection efficiency.

When considering the effect of using different substrates, no significant correlation between measured detection efficiency was found with respect to conductivity, band gap or ionisation energy. The electronic and nuclear stopping powers of ions implanted into substrates were simulated using the Monte Carlo simulation package, SRIM (Stopping Ranges of Ions in Matter \cite{ziegler2010srim}). The stopping powers for the ion species and substrates of interest as a function of implantation energy are plotted in Figure S.2
, however, we found no significant correlation between the stopping powers (or ratios and differences thereof) and the detection efficiencies measured in this work.

Surface roughness has also been observed to have an effect on secondary electron yield variation \cite{baglin2000secondary, seiler1983secondary}, which may explain some results which do not fit the overall trend observed, particularly with respect to some of the samples for which there was a significant difference between the two charge states with the same implantation energy. 

A more advanced theoretical approach will be required in order to obtain a model of secondary electron emission with respect to ion species and substrate parameters and their impact on detection efficiency as measured through channeltron secondary electron detectors. This would potentially provide an indicative model which may be used to estimate the detection efficiency of ions implanted into substrates of technological interest without having to conduct individual detection efficiency experiments.


The high bandwidth detection signal processing capabilities in the P-NAME system makes it possible to incorporate detection efficiency investigations with large field of view STEM imaging with atomic resolution to confirm the accuracy of counting the number of ions into an implantation site, which can enable real-time counting of the number of ions in a rapid and reliable manner for the fabrication of future quantum technologies. 

%% file: Sections/SummaryAndConclusions.tex
\section{Conclusion}

The detection efficiencies presented here provide an insight into the confidence with which a single ion (or cluster) may be implanted. While not all of the ion-substrate combinations are of immediate technological interest, systems which may be employed for deterministic single defect technologies are reported with high detection efficiencies. For silicon, 50 keV P was found to have a detection efficiency of 82 ($\pm$4)\%, and 50 keV (\textsuperscript{121}Sb\textsuperscript{123}Sb) an efficiency of 94 ($\pm$7)\%. When these same ions were implanted into SiO\textsubscript{2}, the detection efficiency saw a marked increase of 90 ($\pm$2)\% and 105 ($\pm$8)\% for the P ion and Sb cluster, respectively. In diamond, the detection efficiency of Ge was found to be 70 ($\pm$2)\%. These results are comparable to those obtained using IBIC detectors of similar energies, only without the requirement for electrodes specifically for the detection mechanism, confirming that deterministic ion implantation using secondary electron emission is a viable route to flexible and scalable qubit array fabrication, particularly with ion species available in liquid metal alloy ion sources, offing a viable route to qubit fabrication en masse.





%% file: Sections/Method.tex
\section{Experimental Methods}
\label{sec:methods}

\subsection{Deterministic Ion Implantation}

The P-NAME ion implantation system (Q-One, Ionoptika) has a maximum anode voltage of 25 kV, with a variety of elements available through a range of liquid metal alloy ion sources (LMAISs). The mass selectivity of the P-NAME system allows for the selection of a specific element from the source alloy with isotopic resolution \cite{adshead2023high}. The kinetic energy of implanted ion species was controlled using the Wien filter to select the ion mass and charge-state for a particular variation of the anode voltage.

Electrostatic plates within the ion beam column were used to provide ion pulsing, enabling the ion beam to be exposed to the substrate in pulses of controlled width. Two high-sensitivity secondary electron detectors within the main sample chamber of the P-NAME system were oppositely biased to attract and detect charged particles: one biased for secondary electrons and negatively charged ions, and the other for positively charged secondary ions. The ion beam current in the sample space was measured prior to, and following, each experiment using a Faraday cup. To mitigate ion source instability impacting the analysis, only data where the pre- and post-experiment ion beam current differed by $<10$\% are included in the work presented here.

Each experiment was configured to implant an array of of ions into 50 x 50 points, with a 1 \textmu m pitch. The ion beam was directed to each point by adjusting the positioning voltages with the beam blanked. Once positioned, the beam blanking voltage was removed for the desired time (defining the ion pulse width) with an analogue to digital converter (ADC) simultaneously recording the incoming signal from the two detectors. The secondary electrons emitted from the sample surface and collected by the detector entered a channeltron portion of the detection mechanism where the electron led to an avalanche of electrons, appreciably amplifying the original signal. The ADC continued to record the signal long after the ion pulse (40 \textmu s in total). The signal for this pulse was analysed and the coordinate registered as either a `pass' or a `fail'. The ion beam was then moved to the next position in the array (with the beam remaining blanked) and the implantation process repeated. The detector gain may be varied through control of the channeltron voltages, with detection efficiencies as a function of gain presented in the Supporting Information, Table S.1
.

Once a full array had been exposed to a single ion pulse the process was then repeated, returning solely to those positions recorded as a `fail'. This sequence was then repeated until all of the points in the array have a recorded `pass', with the number of attempts to complete an entire array being characteristic of the average number of ions per pulse, the detection efficiency, and the secondary electron emission, which is dependent on the ion beam and implanted material in a particular experiment.

In order to reduce the impact of false positives (dark counts), the signal for each pulse was subject to a `detection window'. The beginning of the detection window scales with the ion beam velocity, in order to account for the time taken for an ion to reach the substrate from the electrostatic pulser plates, and then an emitted secondary electron to reach the detector. The width of the window was directly scaled by the pulse width. Detections outside of this window are considered to be dark counts or slow moving ions, which are not considered to correlate with the implantation of an ion in a given pulse and so are not used in the determination of a successful ion implantation detection. The dark-count rate for the P-NAME system is $\sim$8$\times10^{-6}$ per typical \textmu s pulse, and so are considered to not have a significant impact on the work conducted here.

\subsection{Samples and Preparation}

The undoped silicon (University Wafer) was 500 \textmu m thick (with a native oxide layer), with a resistivity of $>$10,1000 \textohm $\cdot$cm, and and orientation of $<$100$>$. The silicon dioxide sample (Alpha Nanotech) was boron-doped (as grown, 1-50 \textohm $\cdot$cm), $<$100$>$ orientation, with a 200 nm thick SiO\textsubscript{2} layer thermally grown. 
The aluminium oxide (University Wafer) was 430 \textmu m thick, C to M plane orientation. The GaAs (MCP Wafer Technology Ltd.) was 500 um thick, undoped and $<$100$>$ orientation. The diamond (Element Six) was CVD-grown electronic grade. The SiC (MSE Supplies) sample was 500 \textmu m thick, semi-insulating 4H.

Prior to loading into the P-NAME system, all samples were subject to a cleaning process of ultrasonication in acetone for 2 minutes, twice, then once for two minutes in IPA, followed by blow-drying with nitrogen gas. The clean samples were fixed to aluminium sample plates using copper adhesive tape (with the exception of the diamond sample, which was adhered using a double-sided adhesive carbon tab). These sample plates were then loaded onto the P-NAME sample platen for the duration of the experiments while the relevant LMAIS sources were in operation, and stored in sample storage boxes otherwise.